\documentclass[prb,aps,twocolumn,amsmath,amssymb,floatfix,superscriptaddress]{revtex4}

\usepackage{graphics,epstopdf}
\usepackage{epsf,graphics,graphicx}
\usepackage{float}
\usepackage{braket}
\usepackage{color}
\definecolor{dred}{rgb}{0.75,0,0}
\usepackage{soul}
\usepackage[colorlinks=true, citecolor=blue, urlcolor=blue]{hyperref}
\usepackage[T1]{fontenc}

\begin{document}

\title{\Large Designing all possible logic gates in phononic lattices: A theoretical study}

\author{Swaraj Biswas}
\email{swaraj.biswas9804\_r@isical.ac.in}
\affiliation{Physics and Applied Mathematics Unit, Indian Statistical Institute, 203 Barrackpore Trunk Road, Kolkata-700 108, India} 

\author{Santanu K. Maiti}
\email{santanu.maiti@isical.ac.in}
\affiliation{Physics and Applied Mathematics Unit, Indian Statistical Institute, 203 Barrackpore Trunk Road, Kolkata-700 108, India} 

%\date{\today}

\begin{abstract}

We propose a scheme for realizing thermal logic gates at the nanoscale using a phononic ring system. Two atomic sites, placed in close 
proximity to the ring, serve as the inputs for two-input logic operations, while a single proximity site is 
employed for single-input logic functionality. The logic output is encoded in the phonon transmission probability, which is calculated 
within the framework of non-equilibrium Green's function formalism. By appropriately tuning the ring-electrode junction configuration, 
all seven standard logic gates, comprising three fundamental and four combinatorial operations, are successfully realized in different 
phonon frequency regimes. Our results suggest that the proposed logic operations remain valid over a broad range of phonon frequencies,
highlighting the generality and reliability of the proposed approach.

\end{abstract}

\maketitle

\section{Introduction}

Logic gates are fundamental and essential elements in digital electronics. For the past few decades, designing logic gates has drawn significant research interest, and people are actively working on it\cite{1,2,3,4,16,17,18,19}. People use electronic circuits containing diodes and transistors\cite{5,6}, considering electrons as information carriers for designing logic gates. Electronic logic gates\cite{7,8,9,10} form the basic building block in present-day devices and computing systems. They consume electrical power and release non-recyclable heat during logical functioning.

There are two major types of electronic logic devices: charge- and spin-based logic devices\cite{7,11,12,13,14,15}. Traditional devices use electric charge to represent binary input states `$0$' and `$1$', where `$0$' and `$1$' represent low charge and high charge, respectively. On the other hand, spin-based logic devices use electron spin (up and down) as binary inputs. Spin logic devices consume less power and generate lower heat compared to charge-based devices. 

There are two types of logic operation: single logic operation and parallel logic operation\cite{1,7,10,19,20,21,22,23}. The single logic operation corresponds to the computing mode, where one specific Boolean operation is performed at a time for all combinations of inputs. In contrast, parallel logic devices execute multiple logic operations simultaneously using the same or different input combinations. Parallel logic devices have higher processing speed and computation efficiency compared to single logic devices.

Modern devices can perform logic operations using heat flow instead of electric current flow, where phonons act as information carriers. Phonon-based devices have emerged as an active research topic and can show potential applications in nanoscale devices\cite{24,25,26,27,28}. They offer waste heat management during information processing and appreciable thermal efficiency, making them promising for future applications. There are different types of thermal devices such as transistors, rectifiers, amplifiers, switches, etc\cite {29,30,31,32}. These devices serve as basic components for thermal computing systems. Thermal transistors use a third terminal to regulate heat flow, analogous to an electronic transistor. On the other hand, rectifiers allow heat to flow in a preferred direction. A thermal switch can turn the heat current on or off in response to an external control. Thermal amplifiers enhance the input heat signal so that a small input current produces a large output heat current.

Like the above-mentioned thermal devices, a few proposals for designing logic gates have been put forward\cite{33,34}, where the Boolean operations are performed using heat signals as inputs, known as thermal logic gates. These logic gates serve as essential elements in phononic computing systems. Designing suitable logic gates is drawing significant attention in recent times, and, focusing on that, in the present work, we propose all possible thermal logic gates using a phononic ring. The ring is coupled to two heat baths (see Fig.~\ref{fig.1}), maintained at two different temperatures, to form a ring nanojunction. Two separate atomic sites are placed in close proximity to the ring, which act as two inputs for our logical operations. Three primary (OR, AND, and NOT) and four combinatorial (NAND, NOR, XOR, and XNOR) logic operations are performed, and for each logical operation, different combinations of the inputs are considered, and for each of these input combinations, the phonon transmission probability is computed, which defines the output. We define the nanojunction using the mass-spring framework and evaluate phonon transmission probability using the non-equilibrium Green's function (NEGF) technique.  Our results are valid for a broad range of input parameters, which suggests that the proposed prescription can be examined in suitable laboratories. At the end, the construction of phononic rings is briefly described, for the sake of completeness of our study. 

The rest of the work is arranged in the following way. In Sec. II we describe the junction setup for designing the logic gates and theoretical formalism for calculating the phonon transmission probability. Numerical results are elaborately discussed in Sec. III. Then we briefly discuss the possible way of constructing our present setup in Sec. IV, and finally, our work is summarized in Sec. V.

\section{Logic gate setup and Theoretical prescription}

\subsection{General logic gate setup}

\begin{figure}[htbp]
\centering
% -------- Row 1 (single image) --------
\includegraphics[width=0.475\textwidth]{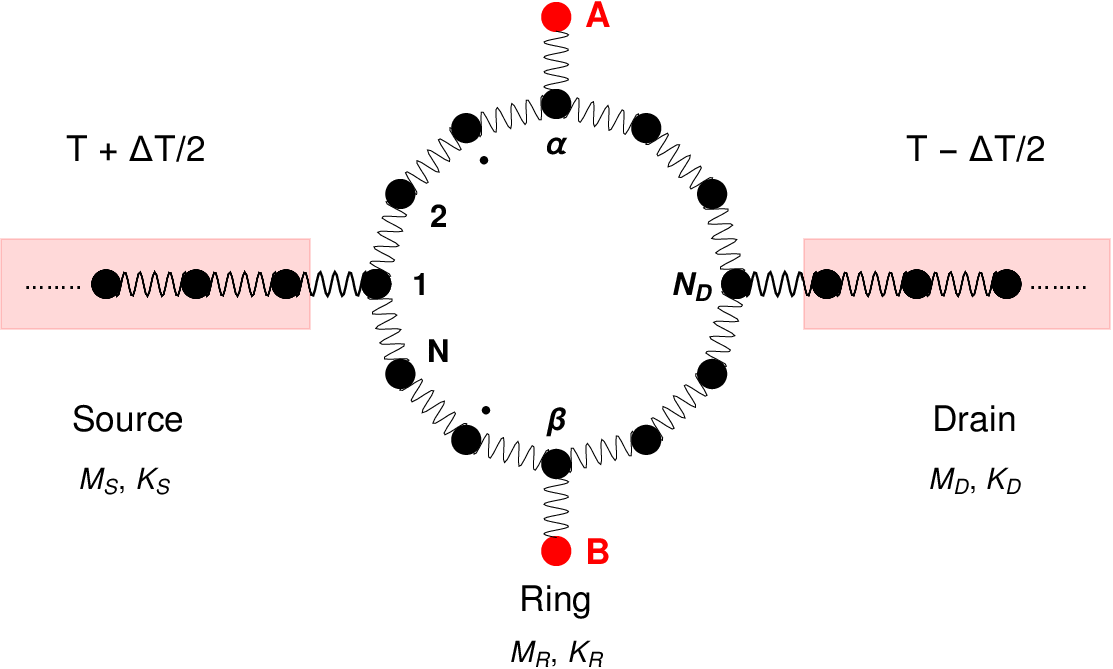}
\caption{(Color online). \small Logic gate setup, where a phononic ring is coupled to source and drain heat baths. Two masses, labeled as $A$ and $B$, are placed in close proximity to the ring that act as logical inputs.}
\label{fig.1}
\end{figure}
Our system comprises a nanoscopic ring containing $N$ number of atomic sites connected to two 1D semi-infinite electrodes, source and drain, shown in Fig.~\ref{fig.1}, where $M_S$ and $K_S$ denote the atomic mass and spring constant, respectively, for the source, $M_D$ and $K_D$ denote the atomic mass and spring constant, respectively, for the drain, and $M_R$ and $K_R$ denote the atomic mass and spring constant, respectively, for the ring. $N_D$ denotes the ring site at which the drain is connected. Two atomic sites, $A$ and $B$, are coupled to specific ring sites $\alpha$ and $\beta$, respectively, which act as inputs for performing the logical operations.  Masses of $A$ and $B$ are denoted as $M_A$ and $M_B$, respectively. For the `ON' state of an input, we have the atomic masses ($M_A$, $M_B$, or both), whereas for the `OFF' state, masses are not present. Spring constants attached to masses $M_A$ and $M_B$ are given as $K_A$ and $K_B$, respectively. 

We assume the source and drain to be identical in nature and set at temperatures $T + \Delta T/2$ and $T -\Delta T/2$, respectively, where $T$ denotes the system's equilibrium temperature and $ \Delta T$ is the temperature difference between the source and the drain. Ring and electrodes are described as phononic spring-mass systems. The two-terminal phonon transmission probability {\Large$\tau_{\scriptscriptstyle ph}$} acts as the output for all logic operations. Method for calculating {\Large$\tau_{\scriptscriptstyle ph}$} is discussed in the next part.

\subsection{Evaluation of phonon transmission probability}

Theoretical prescription for the evaluation of phonon transmission probability {\Large$\tau_{\scriptscriptstyle ph}$} using the NEGF formalism\cite{35,36}, written as, 
\begin{equation}
    \scalebox{1.4}{$\tau_{\scriptscriptstyle ph}$}(\omega) = Tr\left[\Gamma_S\,G_{ph}\,\Gamma_D\, G_{ph}^\dagger\right],
\end{equation}
where $G_{ph}$ represents the phononic Green's function in the ring, given as 
\begin{equation}
    G_{ph} = \left[\mathbb{M}\omega^2-\mathbb{K} - \Sigma_S-\Sigma_D\right]^{-1}.
\end{equation}
Here, $\mathbb{M}$ denotes a diagonal mass matrix and $\mathbb{K}$ is the spring constant matrix. We denote the self-energy matrices for the source and drain by $\Sigma_S$ and $\Sigma_D$, respectively. The thermal broadening matrices $\Gamma_{S}$ and $\Gamma_{D}$ are expressed as $$\Gamma_{S/D} = i\left[\Sigma_{S/D}-\Sigma_{S/D}^\dagger\right].$$\\We calculate the self-energy matrices using the term $$\Sigma_{S/D}^\prime = -K_{av}~exp\left[2i \sin^{-1}\left(\frac{\omega}{\omega_c}\right)\right],$$\\where $K_{av}$ denotes spring constant at the ring-electrode interface, $\omega$ is frequency of phonon incident from the source, and $\omega_c$ denotes the phonon cutoff frequency for the system. All the matrices $\Sigma_S$, $\Sigma_D$, $\mathbb{M}$, and $\mathbb{K}$ have the same dimension.

We calculate the phonon cutoff frequency ($\omega_c$) using the formula $\omega_c$ = 2$\sqrt{K_{av}/M_{av}}$, where $K_{av}$ is obtained by averaging the spring constants of ring and electrode and with $M_{av}$ being the mass, calculated by averaging the atomic masses of the ring and electrode.

\section{Numerical results and discussion}

In what follows, we discuss our essential results. To begin with, we mention all the parameters chosen in this numerical analysis. For the source (drain), we set the atomic mass and spring constant as $M_{S(D)}=4.7\times10^{-26}\,$kg and $K_{S(D)}=16.9\,$N/m, respectively. We consider $M_R=1.2\times10^{-25}\,$kg and $K_R=13.7\,$N/m for the ring. The masses and spring constants for the atomic sites $A$ and $B$ are considered the same as the ring. Here, we have taken an $8\,$-site ring to show our results. The sites $\alpha$, $\beta$, and $N_D$ can vary depending on the logical operations, and their values are mentioned in the respective parts of our analysis. 

First, we elaborately discuss the primary logic gates OR, AND, and NOT one by one. The other gates are discussed in subsequent parts. 
All logic operations are described in terms of phonon transmission probability {\Large$\tau_{\scriptscriptstyle ph}$}, which is the output. 
For a ring, we get high and low values of transmission depending on available phonon channels in the ring and ring-electrode junction 
configuration. In a ring, there are two possible paths: the upper arm and the lower arm. When phonons travel through the ring, they can 
interfere constructively or destructively depending on the difference between the arm lengths and phonon frequency. For constructive 
interference, the phonon transmission probability is high, and for destructive interference, transmission becomes low. These high and 
low outputs, abbreviated as `$H$' and `$L$', respectively, in the Boolean tables, are considered as `ON' and `OFF' states of the output,
respectively, for different logic operations. We refer `$0$'
and `$1$' to represent the `OFF' and `ON' states of an input, respectively. Here, it is relevant to note that `$0$' corresponds to the 
absence of atomic site $A$ or $B$, not the zero mass, and `$1$' denotes the presence of the atomic site $A$ or $B$ with the appropriate mass.  
\begin{figure}[htbp]
\centering
% -------- Row 1 (single image) --------
\includegraphics[width=0.35\textwidth]{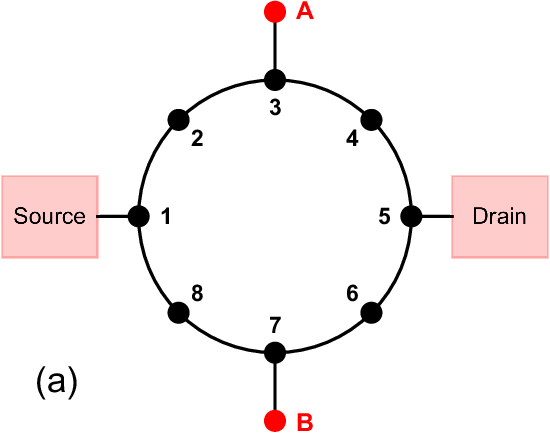}\par\vspace{0.4cm}
% -------- Row 2 --------
    \includegraphics[width=0.35\textwidth]{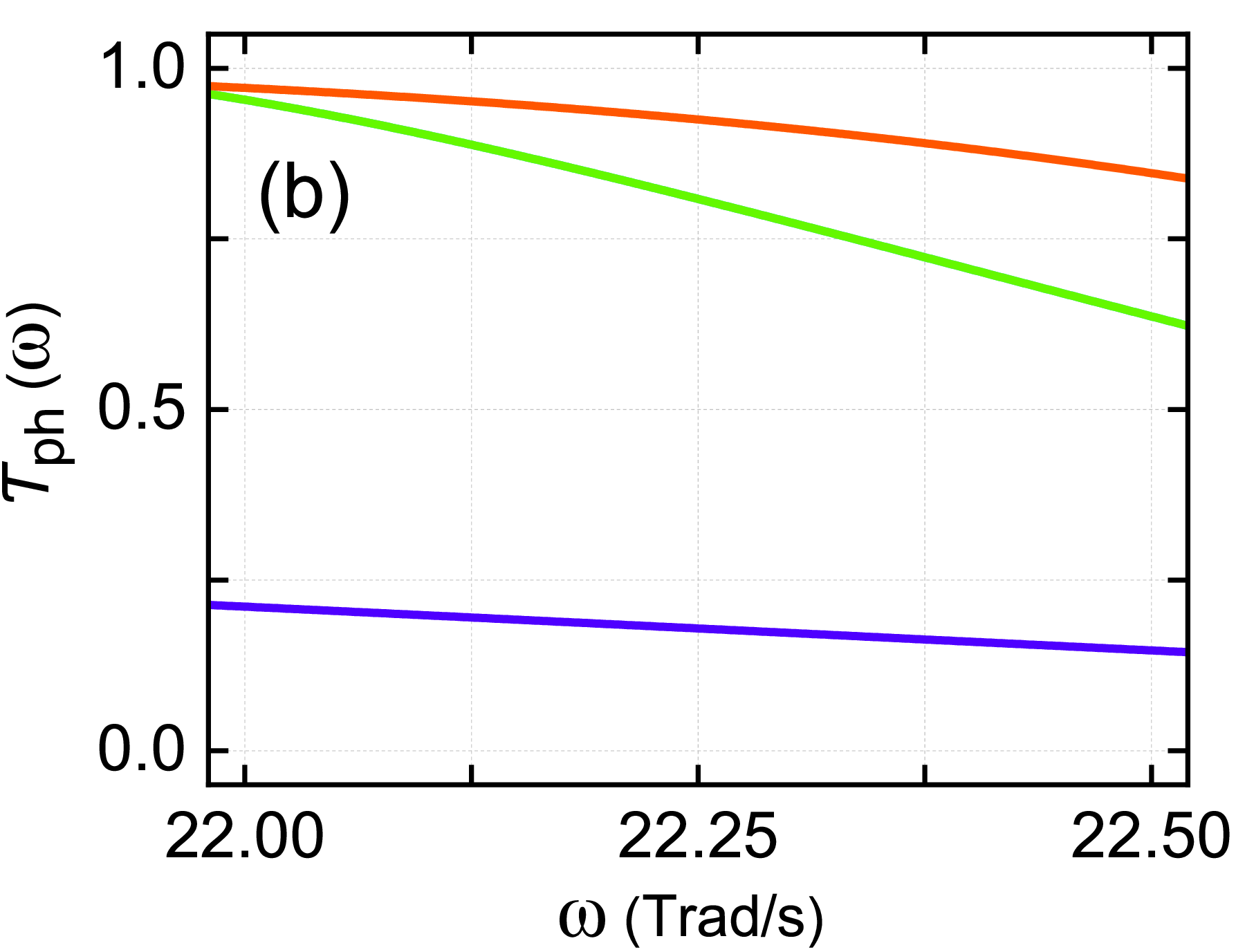}\par\vspace{0.4cm}
\begin{minipage}{0.4\textwidth}
\centering
\renewcommand{\arraystretch}{1.2}
\setlength{\tabcolsep}{8.2pt}
\hspace*{0.62cm}
\begin{tabular}{|c|c|c|}
\hline
\multicolumn{3}{|c|}{Table-$1$}\\
\hline
\shortstack{\rule{0pt}{2ex}Input-$1$ \\($M_A$)} & \shortstack{\rule{0pt}{2ex}Input-$2$ \\ ($M_B$)} & \shortstack{\rule{0pt}{2ex}Output \\(\scalebox{1.4}{$\tau_{ph}$})} \\
\hline
$0$ & $0$ & $0.18$ ($L$) \\
$0$ & $1$ & $0.84$ ($H$) \\
$1$ & $0$ & $0.84$ ($H$) \\
$1$ & $1$ & $0.93$ ($H$) \\
\hline
\end{tabular}
\end{minipage}
\caption{(Color online). \small OR gate operation. (a) Junction setup, where a phononic ring is coupled to two heat baths, source and drain, respectively. Two input sites, labeled as $A$ and $B$, are placed symmetrically to the ring. (b) Phonon transmission probability as a function of phonon frequency under four different input conditions, where the blue line denotes when both inputs are OFF, the green line denotes when only one input is ON, and the orange line shows the case where both inputs are ON. Both for the ($1,0$) and ($0,1$) input conditions, identical transmission probabilities are obtained, and thus, we use one curve for them. Table-$1$ shows the truth table for the OR gate response at $\omega=22.2\,$ Trad/s.}
\label{fig.2}
\end{figure}

\vskip 0.2cm
\noindent 
\textbf{(i)} \textbf{OR gate}: Let us begin with the OR gate operation, which is shown in Fig. \ref{fig.2}. The junction setup 
for the OR gate is shown in Fig. \ref{fig.2}(a), where source and drain are connected symmetrically. Two atomic sites $A$ and $B$ are 
connected at the ring sites $\alpha=3\,$ and $\beta=7\,$, respectively. 

Phonon transmission probability {\Large$\tau_{\scriptscriptstyle ph}$} as a function of phonon frequency $\omega$ is shown 
in Fig.~\ref{fig.2}(b). When two atomic sites are coupled to the ring, they act as local defects, introducing additional vibrational 
modes characterized by their masses and the spring constants connecting them to the ring. When a phonon wave is incident on the ring, 
it interacts with these atomic sites as well as with the parent ring sites. Depending on the ring-electrode junction configuration and 
the presence or absence of the inputs, namely the masses $M_{A}$ and $M_{B}$, the number of available phononic eigenchannels changes, 
resulting in either enhanced or suppressed phonon transmission within the chosen frequency window. For the present setup, the number 
of available channels within the selected frequency range increases when either one or both inputs are ON, whereas it decreases when 
both inputs are OFF. Consequently, the transmission characteristics exhibit the behavior shown by the different colored curves 
in Fig.~\ref{fig.2}(b).

Table-$1\,$ shows the truth table for the OR gate response at a typical frequency $\omega=22.2\,$Trad/s, selected from our chosen window.
\begin{figure}[htbp]
\centering
% -------- Row 1 (single image) --------
\includegraphics[width=0.33\textwidth]{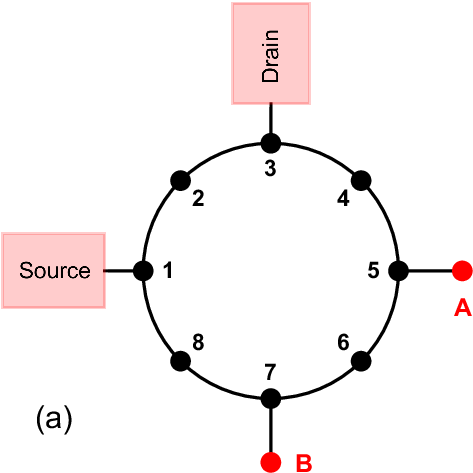}\par\vspace{0.4cm}
% -------- Row 2 --------
    \includegraphics[width=0.35\textwidth]{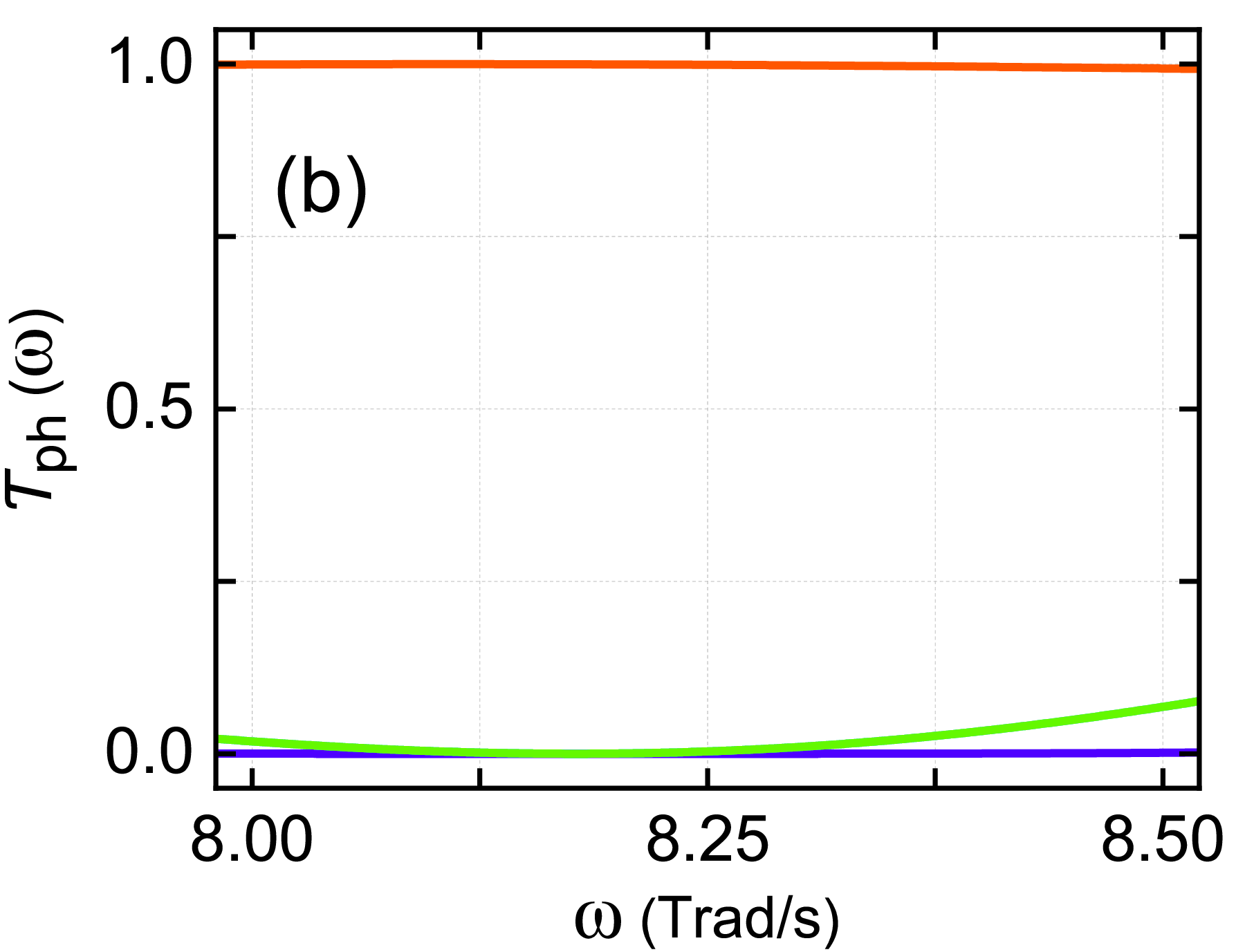}\par\vspace{0.4cm}
\begin{minipage}{0.4\textwidth}
\centering
\renewcommand{\arraystretch}{1.2}
\setlength{\tabcolsep}{8.2pt}
\hspace*{0.73cm}
\begin{tabular}{|c|c|c|}
\hline
\multicolumn{3}{|c|}{Table-$2$} \\
\hline
\shortstack{\rule{0pt}{2ex}Input-$1$ \\($M_A$)} & \shortstack{\rule{0pt}{2ex}Input-$2$ \\ ($M_B$)} & \shortstack{\rule{0pt}{2ex}Output \\(\scalebox{1.4}{$\tau_{ph}$})} \\
\hline
$0$ & $0$ & $0$ ($L$) \\
$0$ & $1$ & $0$ ($L$)\\
$1$ & $0$ & $0$ ($L$)\\
$1$ & $1$ & $0.99$ ($H$)\\
\hline
\end{tabular}
\end{minipage}
\caption{(Color online). \small AND gate operation. (a) Junction setup, where an $8\,$-site ring is connected to the heat baths, source and drain, respectively. Two additional sites, $A$ and $B$, are attached to the lower arm of the ring. (b) Phonon transmission spectrum, where different colors correspond to the same meaning as depicted in Fig.~\ref{fig.2}(b). We use single curve for ($1,0$) and ($0,1$) input conditions, as for both these cases, identical transmission probabilities are obtained. Table-$2$ shows the truth table for the AND gate response at $\omega=8.2\,$ Trad/s.}
\label{fig.3}
\end{figure}

\vskip 0.2cm
\noindent 
\textbf{(ii)} \textbf{AND gate}: AND gate operation is shown in Fig. \ref{fig.3}. Figure \ref{fig.3}(a) shows the junction setup for AND gate operation, where source and drain are asymmetrically connected. The atomic sites $A$ and $B$ are coupled to the ring sites $\alpha=5\,$ and $\beta=7$, respectively.

Now, we discuss the variation of phonon transmission probability for different input conditions.
\begin{figure}[htbp]
\centering
% -------- Row 1 (single image) --------
\includegraphics[width=0.31\textwidth]{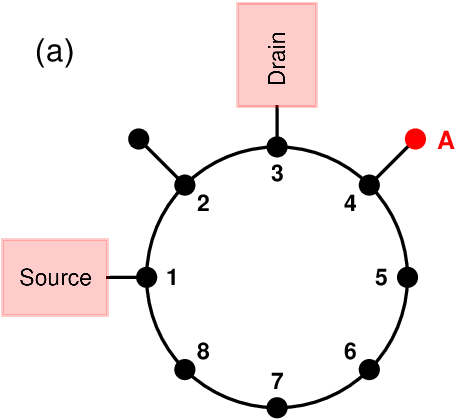}\par\vspace{0.4cm}
% -------- Row 2 --------
    \includegraphics[width=0.35\textwidth]{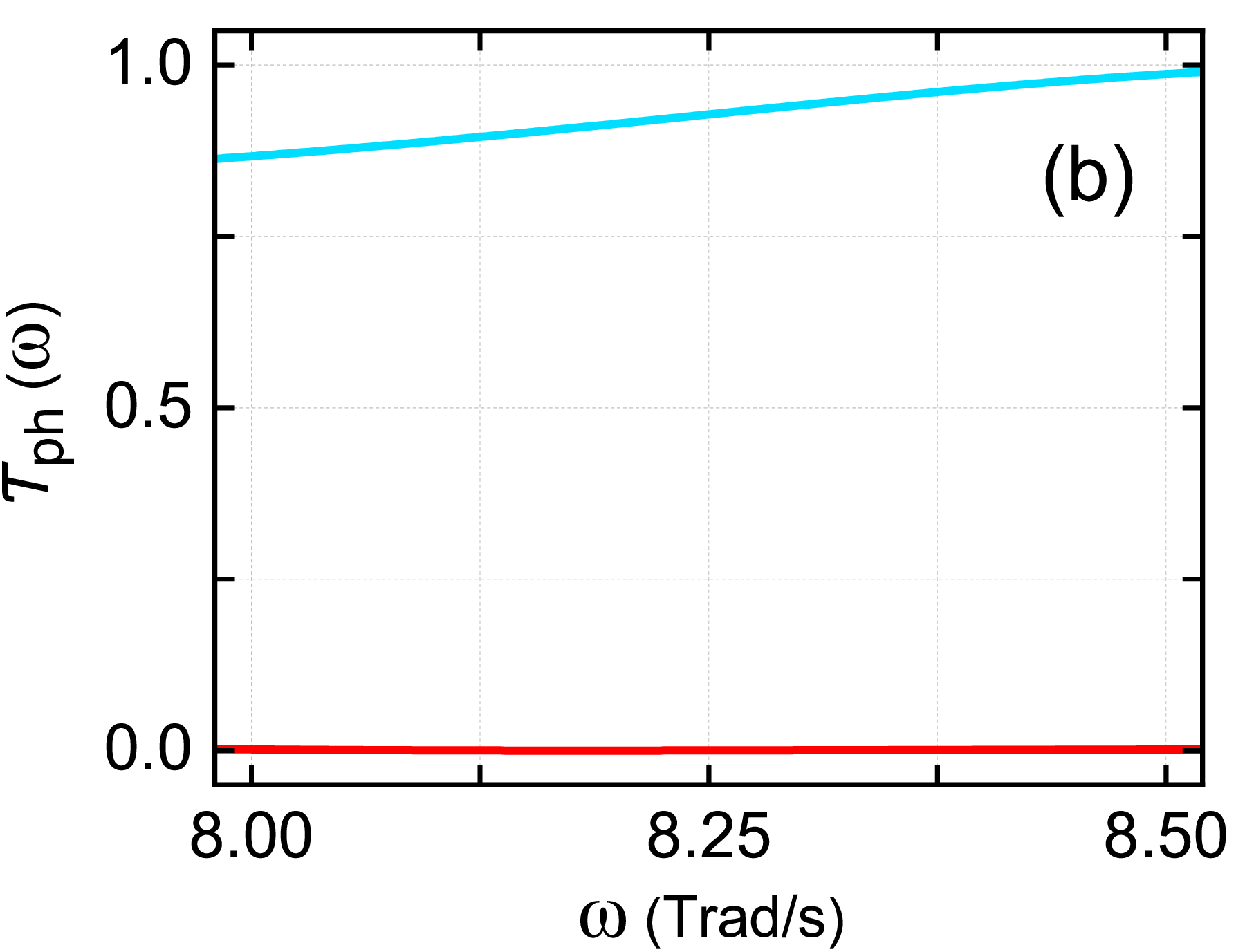}\par\vspace{0.4cm}
\begin{minipage}{0.4\textwidth}
\centering
\renewcommand{\arraystretch}{1.2}
\setlength{\tabcolsep}{9pt}
\hspace*{0.72cm}
\begin{tabular}{|c|c|c|}
\hline
\multicolumn{2}{|c|}{Table-$3$} \\
\hline
Input ($M_A$) & Output (\scalebox{1.4}{$\tau_{ph}$}) \\
\hline
$0$ & $0.98$ ($H$) \\
$1$ & $0$ ($L$) \\
\hline
\end{tabular}
\end{minipage}
\caption{(Color online). \small NOT gate operation. (a) Junction setup, where a ring, with two additional sites, connected with two heat 
baths, source and drain, respectively. Among these two additional sites, the site labeled `A' serves as the input, while the other is part 
of the parent lattice. (b) Phonon transmission probability as a function of phonon frequency, where the cyan line denotes when the input 
is OFF, and the red line indicates the ON state of the input. The truth table for the NOT gate response at $\omega=8.5\,$ Trad/s is shown 
in Table-$3$.}
\label{fig.4}
\end{figure}
In Fig. \ref{fig.3}(b), variation of phonon transmission probability ({\Large$\tau_{\scriptscriptstyle ph}$}) with phonon frequency ($\omega$) is shown. In this case, the atomic sites are connected on the same arm. So, one arm becomes totally inequivalent to the other arm. When phonons propagate through these arms, they experience different environments. In one arm, the absence of any attached mass allows the phonons to propagate freely. For the other arm, the scenario is completely different, where phonons interact with the attached masses and experience scattering as described earlier for the OR gate. Transmission can be high or low depending on constructive or destructive interference between two phonon waves, coming out of the upper and lower arms of the ring, respectively. {\Large$\tau_{\scriptscriptstyle ph}$} is almost zero for all input combinations except one case when both inputs are ON, where transmission probability is nearly $1\,$ for the entire frequency window. 

Table-$2\,$ shows the truth table for the AND gate response at $\omega=8.2\,$Trad/s.

\vskip 0.2cm
\noindent 
\textbf{(iii)} \textbf{NOT gate}: Figure \ref{fig.4} summarizes the NOT gate operation. Junction configuration for NOT gate operation is given in Fig.~\ref{fig.4}(a), where source and drain are asymmetrically connected, where the atomic site $A$ acts as the input for NOT gate operation. 

Phonon transmission probability ({\Large$\tau_{\scriptscriptstyle ph}$}) as a function of phonon frequency ($\omega$) is shown in Fig. \ref{fig.4}(b). Here, when the input is ON, we observe zero phonon transmission probability across the entire frequency range, as shown by the red line. On the other hand, a very high transmission line is obtained as we disconnect $M_A$, shown in cyan. This is due to the quantum interference effect of phonon waves propagating through different arms of the ring.

Table-$3\,$ shows the NOT gate response at $\omega=8.5\,$Trad/s.
\begin{figure}
    \includegraphics[width=0.35\textwidth]{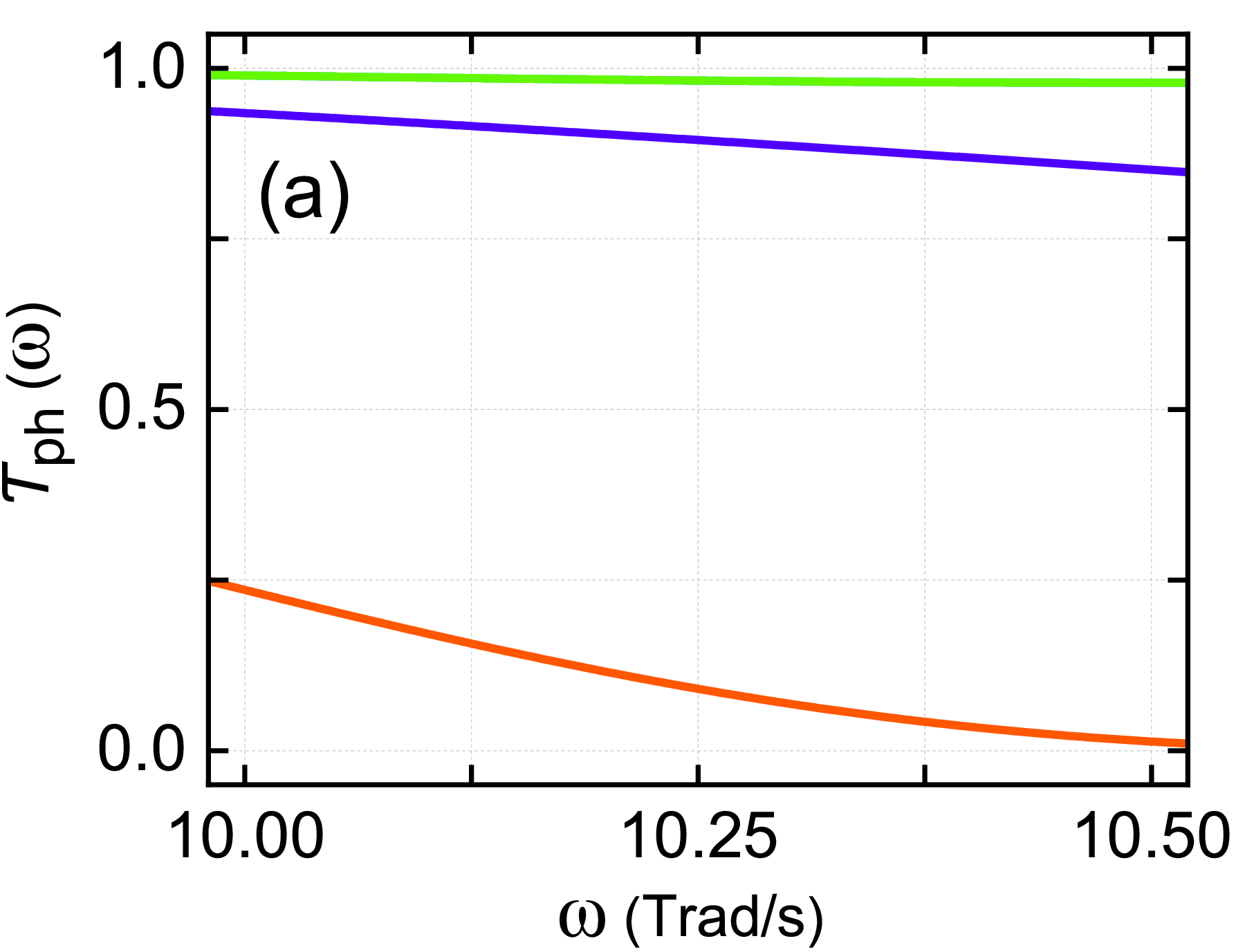}\par\vspace{0.4cm}
\begin{minipage}{0.4\textwidth}
\centering
\renewcommand{\arraystretch}{1.2}
\setlength{\tabcolsep}{8pt}
\hspace*{0.62cm}
\begin{tabular}{|c|c|c|}
\hline
\multicolumn{3}{|c|}{Table-$4$} \\
\hline
\shortstack{\rule{0pt}{2ex}Input-$1$ \\($M_A$)} & \shortstack{\rule{0pt}{2ex}Input-$2$ \\ ($M_B$)} & \shortstack{\rule{0pt}{2ex}Output \\(\scalebox{1.4}{$\tau_{ph}$})} \\
\hline
$0$ & $0$ & $0.85$ ($H$) \\
$0$ & $1$ & $0.97$ ($H$)\\
$1$ & $0$ & $0.97$ ($H$)\\
$1$ & $1$ & $0.01$ ($L$)\\
\hline
\end{tabular}
\end{minipage}
\caption{(Color online). \small NAND gate operation. Junction setup is the same as described in Fig. \ref{fig.2}(a). (a) Phonon transmission probability as a function of phonon frequency, where different colored curves represents the same meaning as described in Fig.~\ref{fig.2}(b). Both for the ($1,0$) and ($0,1$) input conditions, identical transmission probabilities are obtained, and thus, we use one curve for them. Table-$4$ shows the truth table for the NAND gate response at $\omega=10.5$ Trad/s.}
\label{fig.5}
\end{figure}

\vskip 0.2cm
\noindent 
\textbf{(iv)} \textbf{NAND gate}: We show the NAND gate operation in Fig.~\ref{fig.5}. Junction setup for the NAND gate is shown in Fig.~\ref{fig.2}(a). The atomic sites $A$ and $B$ are attached to the ring sites $\alpha=3\,$ and $\beta=7\,$, respectively. 

Phonon transmission function ({\Large$\tau_{\scriptscriptstyle ph}$}) as a function of phonon frequency ($\omega$) for the NAND gate response is shown in Fig.~\ref{fig.5}(a). Here, a suitable frequency window is considered for analyzing the NAND gate response. Transmission is very high in all cases except when masses $M_A$ and $M_B$ are present. A low transmission value is obtained when both inputs are ON. The high and low values of phonon transmission probability are consistent with constructive or destructive interference of two outgoing phonon waves, which depend on the interaction of phonons with local defects, i.e., the masses $M_A$ and $M_B$, as described earlier for previous gates.

The truth table for the NAND gate response at $\omega=10.5\,$Trad/s is shown in Table-$4$.
\begin{figure}
    \includegraphics[width=0.35\textwidth]{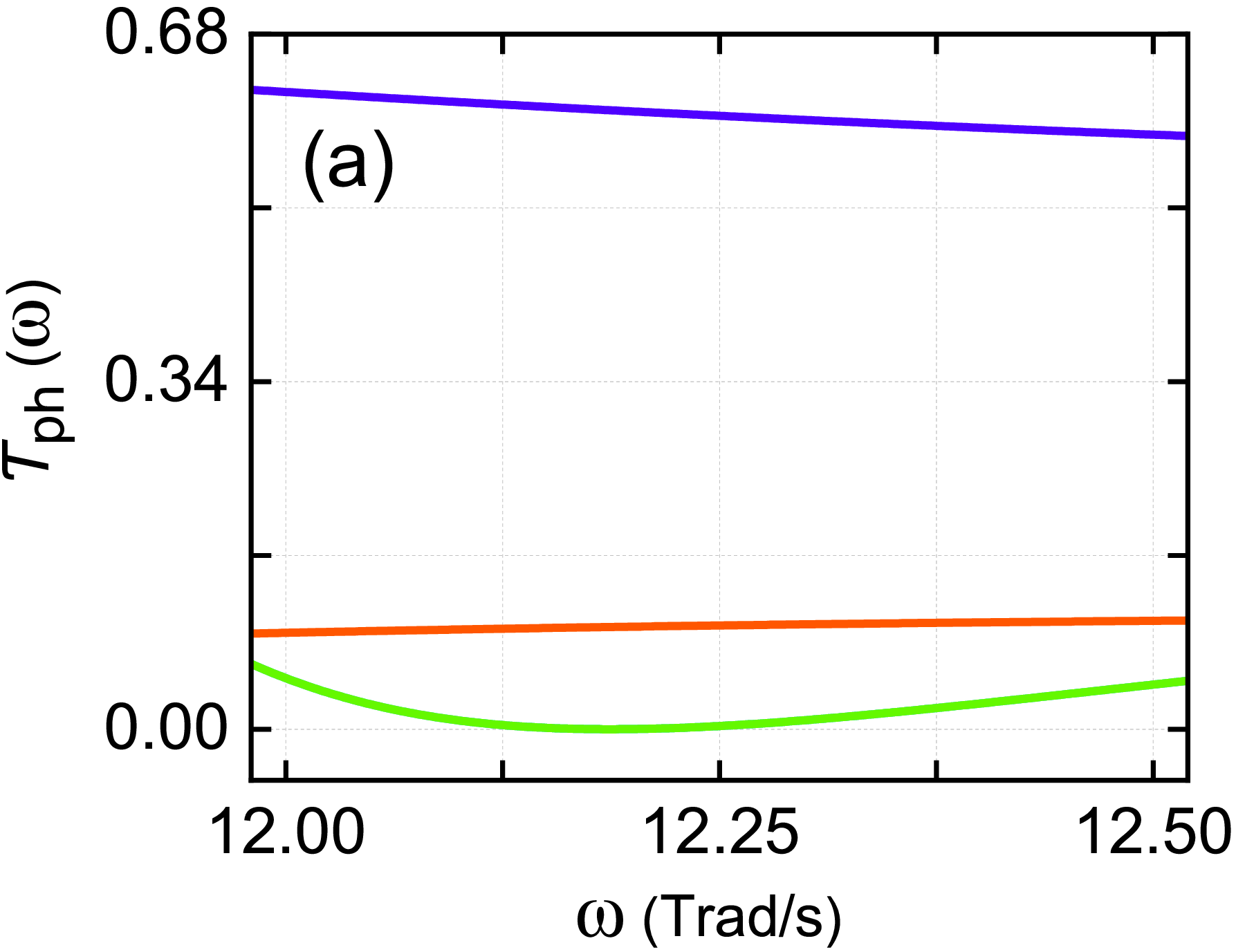}\par\vspace{0.4cm}
\begin{minipage}{0.4\textwidth}
\centering
\renewcommand{\arraystretch}{1.2}
\setlength{\tabcolsep}{6.8pt}
\hspace*{0.84cm}
\begin{tabular}{|c|c|c|}
\hline
\multicolumn{3}{|c|}{Table-$5$} \\
\hline
\shortstack{\rule{0pt}{2ex}Input-$1$ \\($M_A$)} & \shortstack{\rule{0pt}{2ex}Input-$2$ \\ ($M_B$)} & \shortstack{\rule{0pt}{2ex}Output \\(\scalebox{1.4}{$\tau_{ph}$})} \\
\hline
$0$ & $0$ & $0.60$ ($H$) \\
$0$ & $1$ & $0$ ($L$)\\
$1$ & $0$ & $0$ ($L$)\\
$1$ & $1$ & $0.10$ ($L$)\\
\hline
\end{tabular}
\end{minipage}
\caption{(Color online). \small NOR gate operation. Junction setup is the same as described in Fig. \ref{fig.2}(a). (a) Variation of phonon transmission probability with phonon frequency, where different colors correspond to identical meaning as described in Fig. \ref{fig.2}(b). Identical transmission probabilities are obtained for the ($1,0$) and ($0,1$) input conditions, and thus, we use a single curve for them. Table-$5$ shows the truth table for the NOR gate response at $\omega=12.2$ Trad/s.}
\label{fig.6}
\end{figure}

\vskip 0.2cm
\noindent 
\textbf{(v)} \textbf{NOR gate}: The NOR gate operation is summarized in Fig. \ref{fig.6}. Junction setup for the NOR gate operation is shown in Fig. \ref{fig.2}(a), where the atomic sites $A$ and $B$ are coupled to the ring sites $\alpha=3\,$ and $\beta=7\,$, respectively.  

In Fig.~\ref{fig.6}(a), the phonon transmission probability is shown as a function of phonon frequency to analyze the operation of the NOR gate. A small frequency range is considered, which shows the NOR logic operation. When both inputs are OFF, phonons travel without any scattering. So, phonon waves from the different arms interfere constructively in the chosen frequency regime. So, phonon transmission is high in this case. For other input combinations, we get a low value of phonon transmission.

The truth table for the NOR gate response at $\omega=12.2\,$Trad/s is shown in Table-$5$.
\begin{figure}
\centering
% -------- Row 1 (single image) --------
\includegraphics[width=0.31\textwidth]{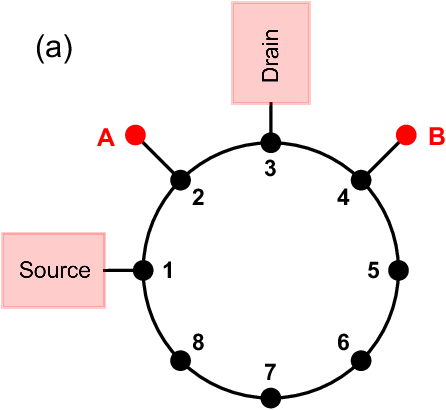}\par\vspace{0.4cm}
\includegraphics[width=0.35\textwidth]{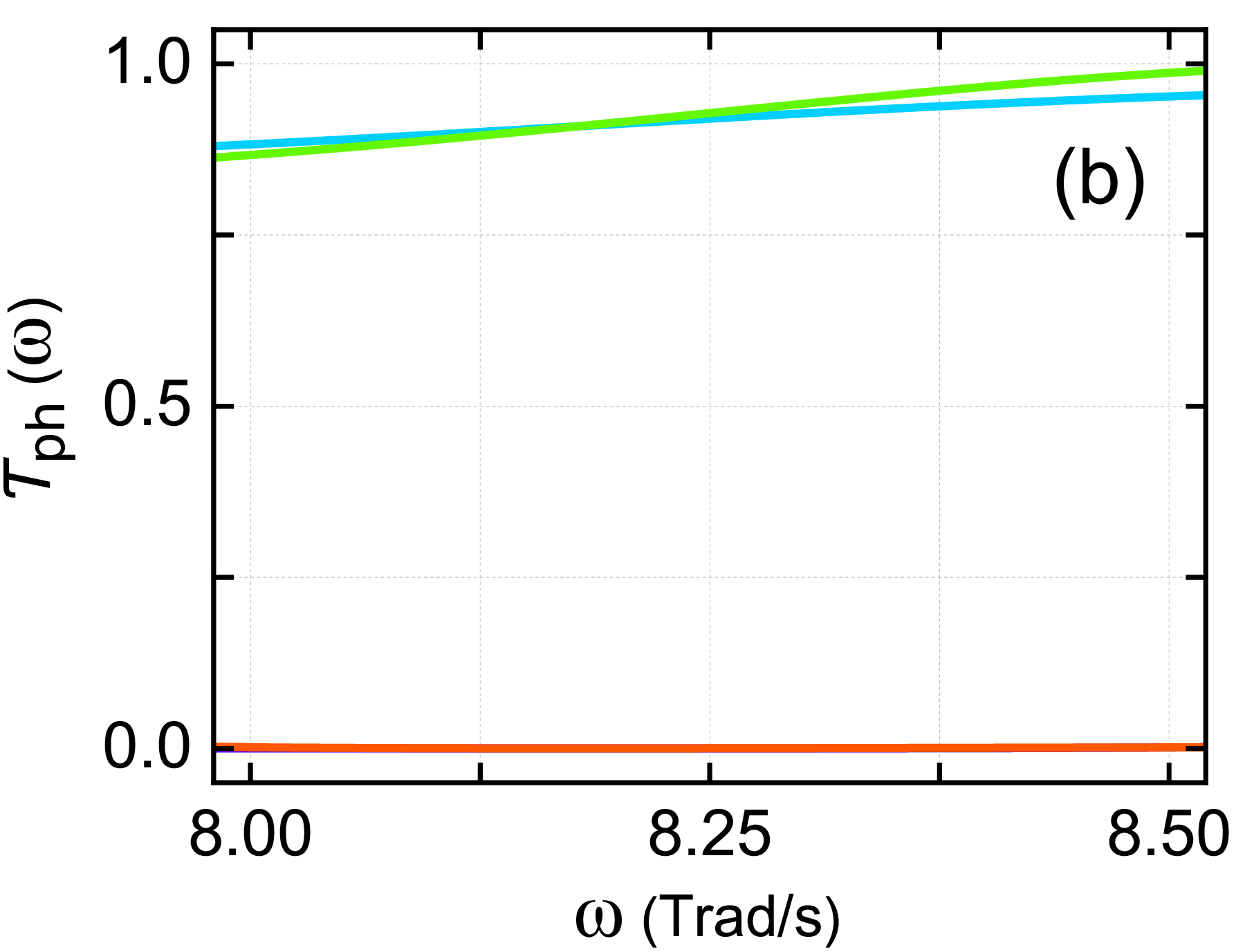}\par\vspace{0.4cm}
\begin{minipage}{0.4\textwidth}
\centering
\renewcommand{\arraystretch}{1.2}
\setlength{\tabcolsep}{8pt}
\hspace*{0.76cm}
\begin{tabular}{|c|c|c|}
\hline
\multicolumn{3}{|c|}{Table-$6$} \\
\hline
\shortstack{\rule{0pt}{2ex}Input-$1$ \\($M_A$)} & \shortstack{\rule{0pt}{2ex}Input-$2$ \\ ($M_B$)} & \shortstack{\rule{0pt}{2ex}Output \\(\scalebox{1.4}{$\tau_{ph}$})} \\
\hline
$0$ & $0$ & $0$ ($L$)\\
$0$ & $1$ & $0.95$ ($H$)\\
$1$ & $0$ & $0.98$ ($H$) \\
$1$ & $1$ & $0$ ($L$) \\
\hline
\end{tabular}
\end{minipage}
\caption{(Color online). \small XOR gate operation. (a) Schematic diagram of a phononic ring connected to two heat baths, source and drain, respectively. Two atomic sites $A$ and $B$ are coupled to the upper and lower arms of the ring, respectively. (b) Phonon transmission probability as a function of phonon frequency, where the red line denotes when both inputs are either ON or OFF, the cyan line denotes when input-$2$ is ON, and the green line denotes when input-$1$ is ON. Table-$6$ shows the truth table for the XOR gate response at $\omega=8.5$ Trad/s.}
\label{fig.7}
\end{figure}

\vskip 0.2cm
\noindent 
\textbf{(vi)} \textbf{XOR gate}: The XOR gate operation is shown in Fig.~\ref{fig.7}. Figure \ref{fig.7}(a) shows the junction setup for the XOR logic operation, where the atomic sites $A$ and $B$ are connected to the ring sites $\alpha=2\,$ and $\beta=4\,$, respectively.

Phonon transmission function ({\Large$\tau_{\scriptscriptstyle ph}$}) as a function of phonon frequency ($\omega$) for the XOR gate response is shown in Fig. \ref{fig.7}(b). Transmission is very high when only one of the two attached masses is present. On the other hand, transmission is zero in the whole frequency window when both inputs are in the same state. The high and low values of phonon transmission probability are consistent with the interference effect of the phonon waves, traversing through two arms of the ring.

The truth table for the XOR gate response at $\omega=8.5\,$Trad/s is shown in Table-$6$.

\begin{figure}[ht]
\includegraphics[width=0.35\textwidth]{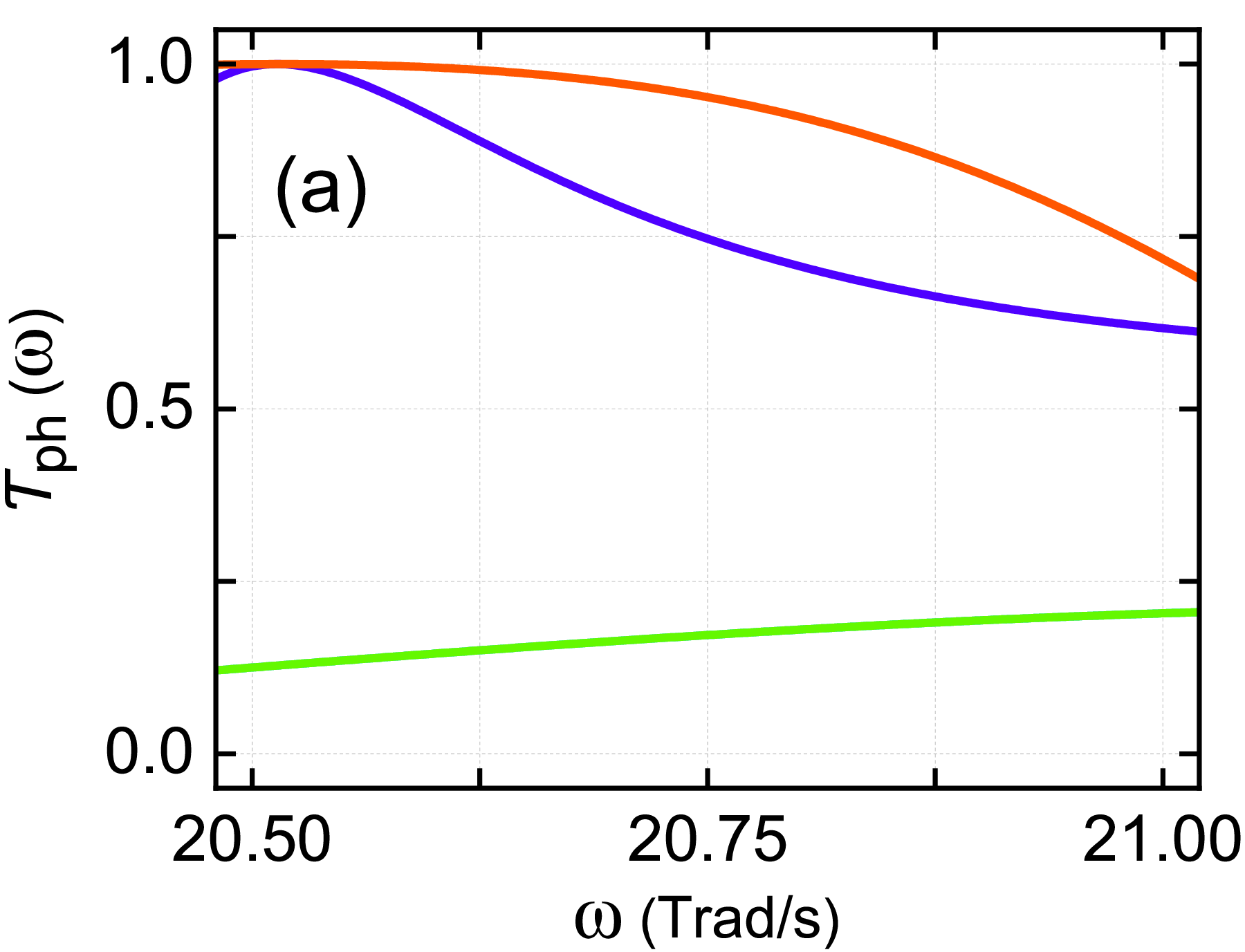}\par\vspace{0.4cm}
\begin{minipage}{0.4\textwidth}
\centering
\renewcommand{\arraystretch}{1.2}
\setlength{\tabcolsep}{8pt}
\hspace*{0.76cm}
\begin{tabular}{|c|c|c|}
\hline
\multicolumn{3}{|c|}{Table-$7$} \\
\hline
\shortstack{\rule{0pt}{2ex}Input-$1$ \\($M_A$)} & \shortstack{\rule{0pt}{2ex}Input-$2$ \\ ($M_B$)} & \shortstack{\rule{0pt}{2ex}Output \\(\scalebox{1.4}{$\tau_{ph}$})} \\
\hline
$0$ & $0$ & $0.99$ ($H$) \\
$0$ & $1$ & $0.12$ ($L$) \\
$1$ & $0$ & $0.12$ ($L$)\\
$1$ & $1$ & $0.99$ ($H$)\\
\hline
\end{tabular}
\end{minipage}
\caption{(Color online). \small XNOR gate operation. Junction setup is the same as depicted in Fig. \ref{fig.3}(a). (a) Variation of phonon transmission probability with phonon frequency, where different colors correspond to identical meaning as described in Fig. \ref{fig.2}(b). For input combinations ($1,0$) and ($0,1$), the same transmission profiles are obtained, and so, the same color is used to denote them. Table-$7$ shows the truth table for the XNOR gate response at $\omega=20.5$ Trad/s.}
\label{fig.8}
\end{figure}

\vskip 0.2cm
\noindent 
\textbf{(vii)} \textbf{XNOR gate}: Now, we finally summarize the XNOR gate operation in Fig. \ref{fig.8}. Junction setup for the XNOR gate 
operation is shown in Fig. \ref{fig.3}(a), where the atomic sites $A$ and $B$ are coupled to the ring sites $\alpha=5$ and $\beta=7$, respectively.

In Fig. \ref{fig.8}(a), the phonon transmission probability is shown as a function of phonon frequency to analyze the operation of the 
XNOR gate. A suitable frequency range is selected that demonstrates the XNOR logic operation. When only one input is ON, phonons 
experience different
environments as they propagate through the two arms. The same scenario is observed in both these cases, where either input is ON, and hence the phonon transmission probability is identical, as denoted by the green line. On the other hand, transmission is high for the other two input combinations, which accurately reflects the operation of the XNOR logic gate.

The truth table for the XNOR gate response at $\omega=20.5\,$Trad/s is shown in Table-$7$.

\section{Experimental perspective}

Thermal logic gates have been experimentally realized in various systems, controlling heat flow rather than electric current~\cite{37,38}.
Our proposed junction setup can be designed by constructing a nanoring, which can be fabricated using electron-beam lithography or some 
other available sophisticated technologies. Two small metallic nanoparticles that act as local resonators (inputs) can be coupled to the
ring using suitable techniques. The system, i.e., the ring with attached particles, should then be connected to two heat reservoirs: one
hot and the other cold. To stay in the linear transport regime, the temperature difference between the two heat baths should be very 
small. The high- and low-output states can be measured using appropriate equipment. Although practical implementation may 
face some challenges due to fabrication imperfections and environmental effects, the rapid progress in phononic and thermal nanodevice
technologies suggests that the experimental realization of the proposed phonon-based logic architecture should be achievable in the 
foreseeable future.

\section{Concluding remarks}

To conclude, in this work, we propose a possible route for designing phonon-based logic gates at the nanoscale. We consider a phononic 
nanoring coupled to two heat baths, one hot and the other cold. The ring and the heat baths are described within a spring-mass framework. 
For two-input logic operations, two atomic sites placed in close proximity to the ring serve as inputs, whereas a single proximity site 
is employed for the one-input logic gate. The output response of each logic operation is quantified through the phonon transmission probability, calculated using the standard non-equilibrium Green's function formalism.

Three fundamental and four combinational logic operations are analyzed in detail. Four different ring-electrode junction configurations 
are utilized to realize the various logic gates. Specifically, OR, NAND, and NOR operations are achieved within the same junction setup, 
while NOT and XOR gates require different configurations. Similarly, AND and XNOR gates are implemented using another common setup. All 
key results remain valid over broad phonon frequency ranges, as confirmed through extensive numerical analysis. We have also verified 
the robustness of our findings for different ring sizes and obtained results consistent with those presented here, and therefore, they 
are not repeated for the sake of brevity. We believe that our findings may provide a viable pathway toward the experimental realization 
of phononic logic devices at the nanoscale.

\end{document}